\documentstyle[12pt,epsfig]{article}
\title{\hspace*{10cm} {\large Budker INP 98-02 \\
\hspace{10cm}hep-ex/9802003} \\
\vspace{0.5cm} 
 {\bf Physics goals and parameters of photon colliders}\thanks{Talk at 
the 2nd Int.Workshop on 
Electron--Electron Interaction at TeV Energies,
 Santa Cruz, CA, USA, September 22--24, 1997. To be published in Int. J. Mod.
Phys. A}}

\author{Valery Telnov\thanks{Email: telnov@inp.nsk.su} \\
\it{Institute of Nuclear Physics,
630090, Novosibirsk, Russia}} 
\date{}

\begin{document}

\newcommand{\EP}{\mbox{e$^+$}}
\newcommand{\EM}{\mbox{e$^-$}}
\newcommand{\EPEM}{\mbox{e$^+$e$^-$}}
\newcommand{\EMEM}{\mbox{e$^-$e$^-$}}
\newcommand{\GG}{\mbox{$\gamma\gamma$}}
\newcommand{\GE}{\mbox{$\gamma$e}}
\newcommand{\GP}{\mbox{$\gamma$e$^+$}}
\newcommand{\TEV}{\mbox{TeV}}
\newcommand{\GEV}{\mbox{GeV}}
\newcommand{\LGG}{\mbox{$L_{\gamma\gamma}$}}
\newcommand{\LEE}{\mbox{$L_{ee}$}}
\newcommand{\WGG}{\mbox{$W_{\gamma\gamma}$}}
\newcommand{\EV}{\mbox{eV}}
\newcommand{\CM}{\mbox{cm}}
\newcommand{\MM}{\mbox{mm}}
\newcommand{\NM}{\mbox{nm}}
\newcommand{\MKM}{\mbox{$\mu$m}}
\newcommand{\SEC}{\mbox{s}}
\newcommand{\CMS}{\mbox{cm$^{-2}$s$^{-1}$}}
\newcommand{\MRAD}{\mbox{mrad}}
\newcommand{\IND}{\hspace*{\parindent}}
\newcommand{\E}{\mbox{$\epsilon$}}
\newcommand{\EN}{\mbox{$\epsilon_n$}}
\newcommand{\EI}{\mbox{$\epsilon_i$}}
\newcommand{\ENI}{\mbox{$\epsilon_{ni}$}}
\newcommand{\ENX}{\mbox{$\epsilon_{nx}$}}
\newcommand{\ENY}{\mbox{$\epsilon_{ny}$}}
\newcommand{\EX}{\mbox{$\epsilon_x$}}
\newcommand{\EY}{\mbox{$\epsilon_y$}}
\newcommand{\BI}{\mbox{$\beta_i$}}
\newcommand{\BX}{\mbox{$\beta_x$}}
\newcommand{\BY}{\mbox{$\beta_y$}}
\newcommand{\SX}{\mbox{$\sigma_x$}}
\newcommand{\SY}{\mbox{$\sigma_y$}}
\newcommand{\SZ}{\mbox{$\sigma_z$}}
\newcommand{\SI}{\mbox{$\sigma_i$}}
\newcommand{\SIP}{\mbox{$\sigma_i^{\prime}$}}

\maketitle

\begin{abstract} 
Linear colliders offer a unique possibility to study
$\gamma\gamma$ and $\gamma$e interactions at the energies
0.1--2 TeV. This option is
now included  in design reports of NLC, JLC and TESLA/SBLC. This paper
includes: status of photon colliders, new possibilities in study of
Higgs boson, ways to achieve  high luminosities. 
\end{abstract}

\section{Introduction}

\IND\ It is very likely that linear colliders with the c.m.s energies of
0.2--2 TeV will be built sometime, may be in about ten years from
now~\cite{LOEW}. Besides \EPEM collisions, linear colliders give an
unique possibility to study \GG\ and \GE\ interactions at energies
and luminosities comparable to those in \EPEM\
collisions~\cite{GKST81}-\cite{GKST84}.

\begin{figure}[!hbt]
\hspace*{1cm}\begin{minipage}[b]{0.45\linewidth}
\centering
\vspace*{-0.0cm} 
\hspace*{.0cm} \epsfig{file=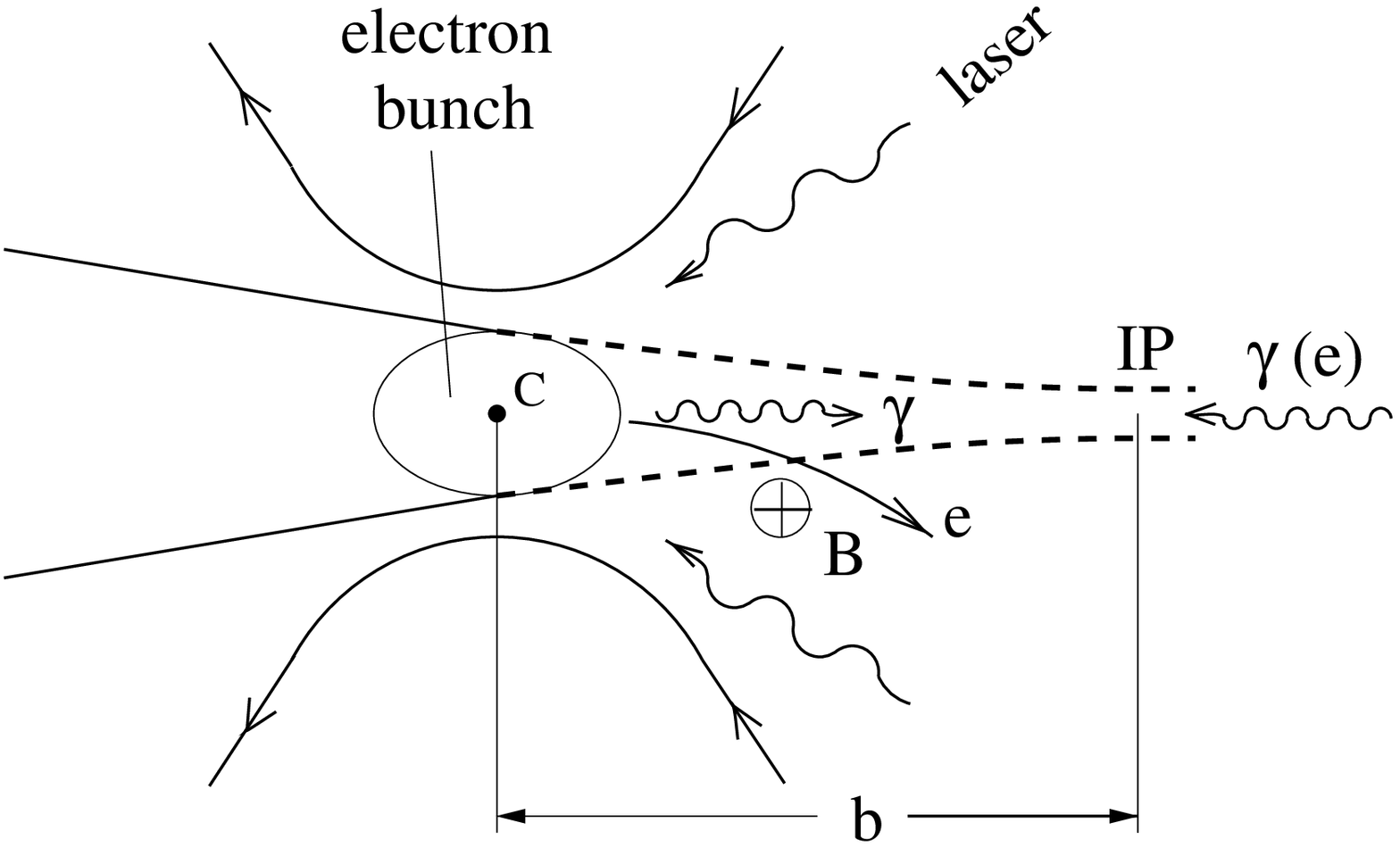,height=6.5cm,angle=-90} 
\vspace*{0.3cm}
\caption{Scheme of  \GG; \GE\ collider.}
\label{ris1}
\end{minipage}%
\hspace*{0.3cm} \begin{minipage}[b]{0.45\linewidth}
\centering
\vspace*{-0.0cm} 
\hspace*{0.5cm} \epsfig{file=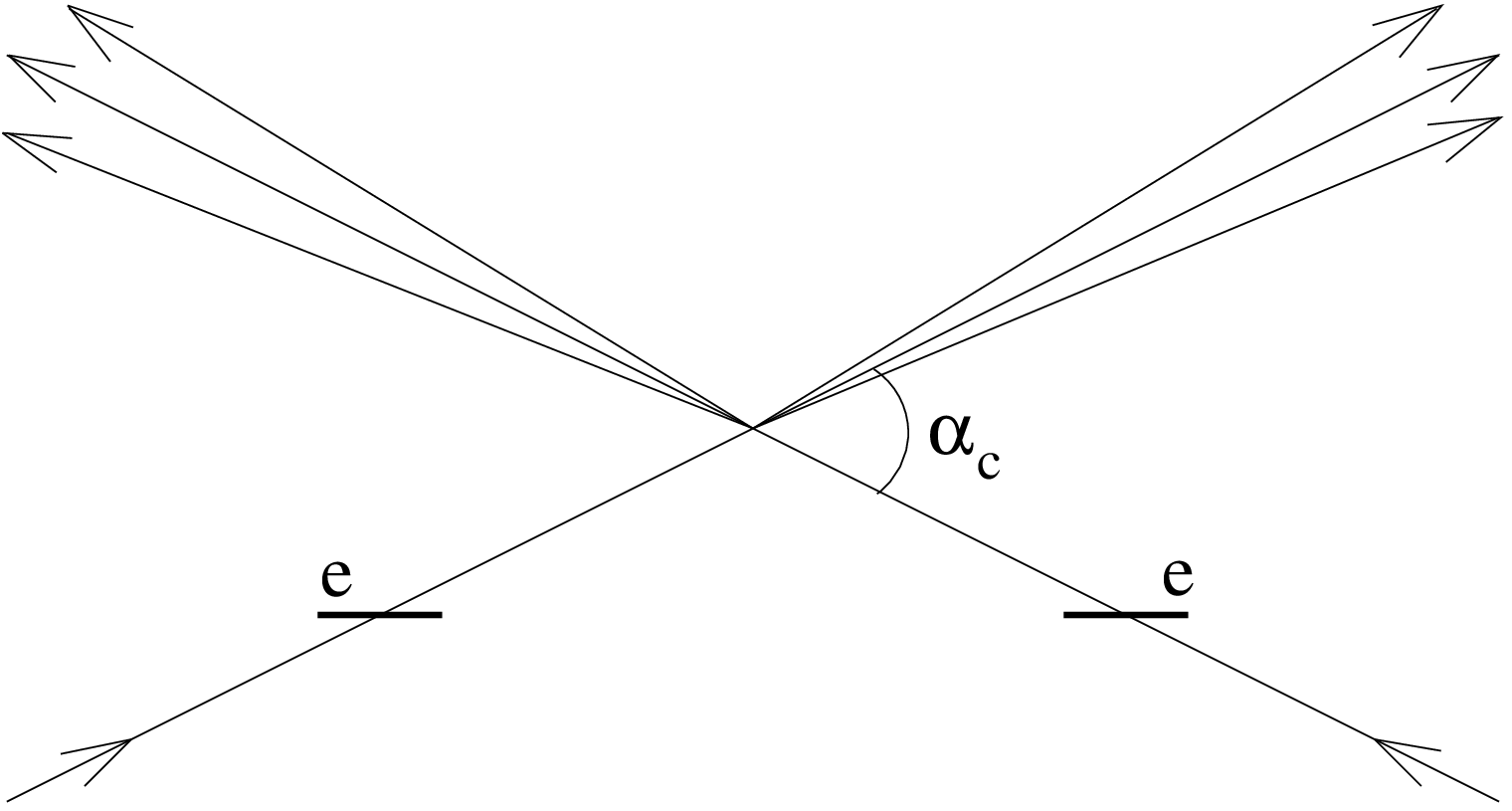,width=0.8in,angle=-90}
\vspace*{1.5cm}
\caption{Crab-crossing scheme}
\label{fig9}
\end{minipage}
\end{figure}

The basic scheme of a photon collider is shown in
fig.~\ref{ris1}. 
Two electron beams
after the final focus system are traveling toward the interaction
point (IP).  At a distance of about 0.1--1 cm upstream from the IP, at
the conversion point (C), the laser beam is focused and Compton
backscattered by electrons, resulting in the high energy beam of
photons.  With reasonable laser parameters one can ``convert'' most of
the electrons into high energy photons. The photon beam follows the
original electron direction of motion with a small angular spread of
the order $1/\gamma $, arriving at the IP  tightly focused, where it
collides with a  similar opposing high energy photon beam or with an
electron beam. The photon spot size at the IP may be almost equal to
that of electrons at IP, and, therefore, the luminosity of \GG, \GE\
collisions will be of the same order of magnitude as the ``geometric''
luminosity of the basic $ee$ beams.  The detailed discussion of photon
colliders  can be found in refs~\cite{GKST83}-\cite{TEL95}.
and in the Berkeley Workshop Proceedings~\cite{BERK}.

  As of today, this option is included into the Conceptual Design Reports of
 NLC~\cite{NLC}, TESLA--SBLC~\cite{TESLA}, and JLC~\cite{JLC} linear
colliders. All these  projects foresee the second
interaction regions for \GG, \GE\ collisions.


   This paper covers three topics: 
\begin{itemize}
\vspace*{-0.3cm}\item  physics at photon
colliders,  paticularly some important remarks on new possibilities
to study the Higgs boson in \GG\ collisions; 
\vspace*{-0.3cm}\item  parameters of photon colliders in `Zero design'
projects of linear colliders; 
\vspace{-0.3cm}\item  ways to achieve very high luminosities.
\end{itemize}

\section{Physics}
\subsection{General remarks}
  The physics in high energy \GG, \GE\ colliders is very rich. The
total number of papers devoted to this subject exceeds one
thousand. Recent reviews of physics at photon colliders can be found,
for instance, in TESLA/SBLC Conceptual Design Report~\cite{TESLA}
(with many references therein) and in the J.Jikia's talk at
``Photon 97''~\cite{JIKIA97}.  In this paper, besides some general words
on the physics in \GG, \GE\ collisions, I would like to discuss some new
important aspects of the Higgs study in \GG\ collisions.
   
\begin{figure}[thb]
\centering
\vspace*{-1.4cm}
\epsfig{file=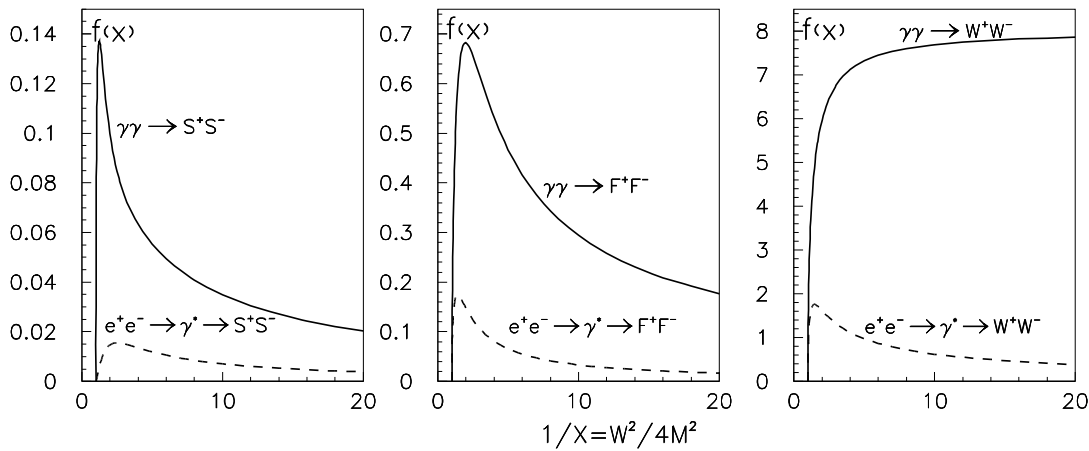,width=6.5in}
\vspace*{-1.2cm}
\caption{ Comparison of  cross sections for charged pair production
in \EPEM\ and \GG\ collisions. The cross section  $\sigma
=(\pi\alpha^2/M^2)f(x)$, P=S (scalars), F (fermions),
W (W-bosons); M is particle mass, $x=W_{p\bar{p}}^2/4M^2$. The functions
$f(x)$ are shown.}

\vspace{-0.3cm}

\label{fig16}
\end{figure}

   For scientific (and political) motivation of building photon colliders the
   following short list of  physics goals can be presented:

\vspace*{-1mm}

\begin{enumerate}
   
\vspace*{-1mm}
 \item Some phenomena can be better studied at photon colliders
   than with pp or \EPEM\ collisions, for example, the
  measurement of the two-photon decay width of the Higgs boson.  Due the
  loop diagrams, all massive  (even ultra-heavy) charged particles
  contribute to this width if their mass is originated by the Higgs
  mechanism.  Some Higgs decay modes and its mass can be
  measured at \GG\ colliders better than in \EPEM\ or pp collisions
  (see the next section).

\vspace*{-1mm} 
\item Cross sections for production of charged scalar,
  lepton and top pairs in \GG\ collisions are larger than those in
  \EPEM\ collisions approximately by a factor of 5 (see
  fig.~ \ref{fig16}); for WW production, this factor is even larger,
  about 10--20.

\vspace*{-1mm} 
\item In \GE\ collisions, charged supersymmetric
  particles with masses higher than in \EPEM\ collisions can be
  produced (a heavy charged particle plus a light neutral), \GG\
  collisions also provide higher accessible masses for particles which
  are produced as a single resonance in \GG\ collisions (such as the Higgs).
\end{enumerate}

These examples together with the fact that the luminosity in \GG\
 collisions is potentially higher than that in \EPEM\ collisions (see
 sect. 4) are very strong arguments in favor of photon colliders.

  A typical distribution of \GG\ or \GE\  luminosity on the  invariant mass
has  high energy peaks at the maximum masses with widths
$\Delta W_{\GG} / W_{\GG} \approx 0.15$,  $\Delta
 W_{\GE}/W_{\GE} \approx 0.05$\cite{TEL95,TESLA}. Below  
the high energy peak, there is usually a flat part of the luminosity
distribution with 1--10 times larger total luminosity (depending
on details of the collision scheme). 

Statistics equal to that in \EPEM\ collisons, can be achieved in \GG\
collisions with much smaller (at least by a factor of five) luminosity. Of
course, in any case data in \GG\ and \EPEM\ collisions are
complimentary to each other because  the coresponding diagrams are different).


\subsection{Higgs in {\boldmath$\gamma\gamma$} collisions}
\begin{figure}[thb]
\hspace*{1cm}\begin{minipage}[b]{0.45\linewidth}
\centering
\hspace*{-2.0cm}
\scalebox{0.9}[1.]{
\includegraphics[width=3.5in, angle=0, bb=90 140 490 425,clip] {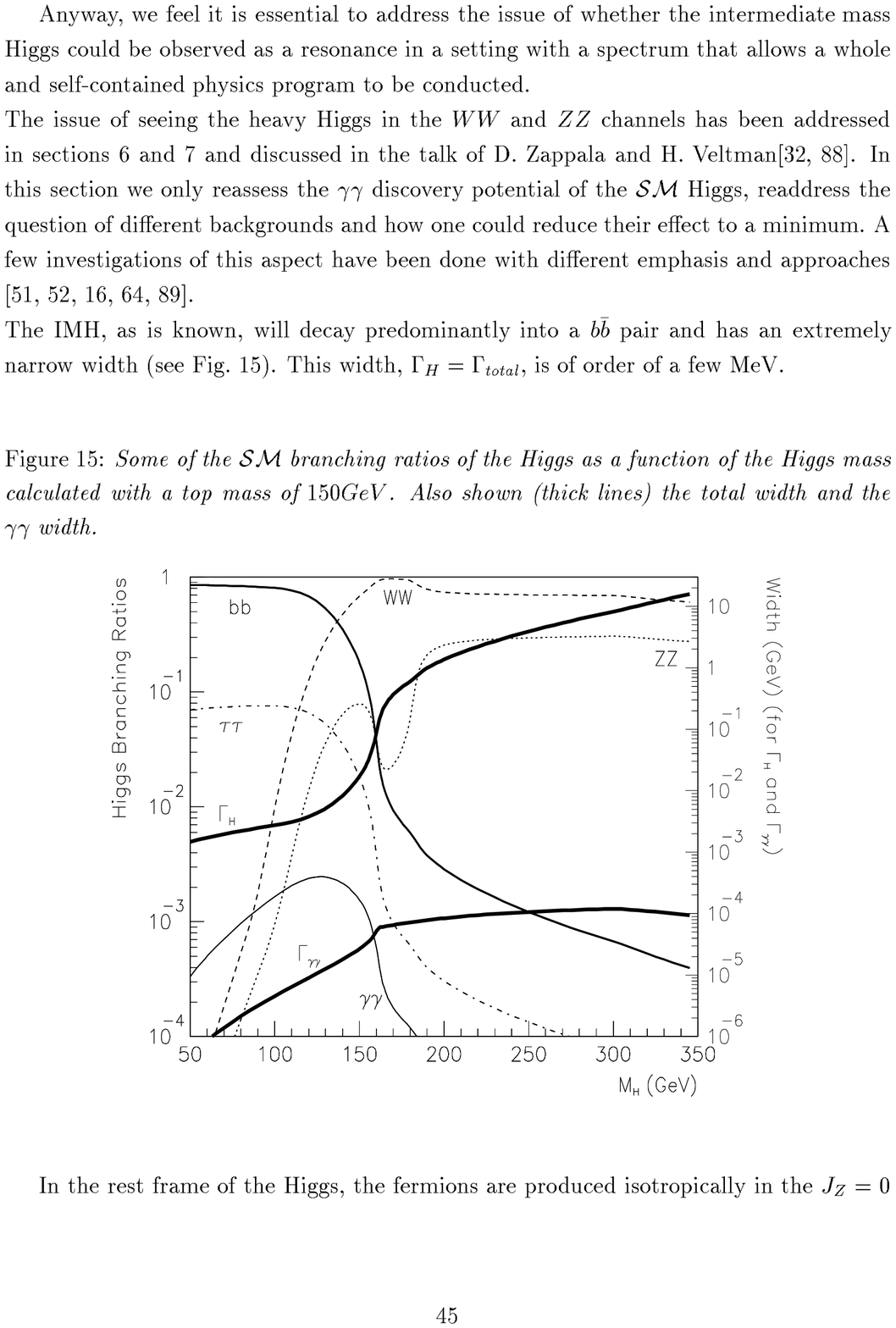}}
\end{minipage}%
\hspace*{0.5cm} \vspace*{0.3cm} \begin{minipage}[b]{0.45\linewidth}
\centering
\scalebox{1.}[1.]{ 
\includegraphics[width=2.43in, angle=0, bb=308 293 457 441,clip] {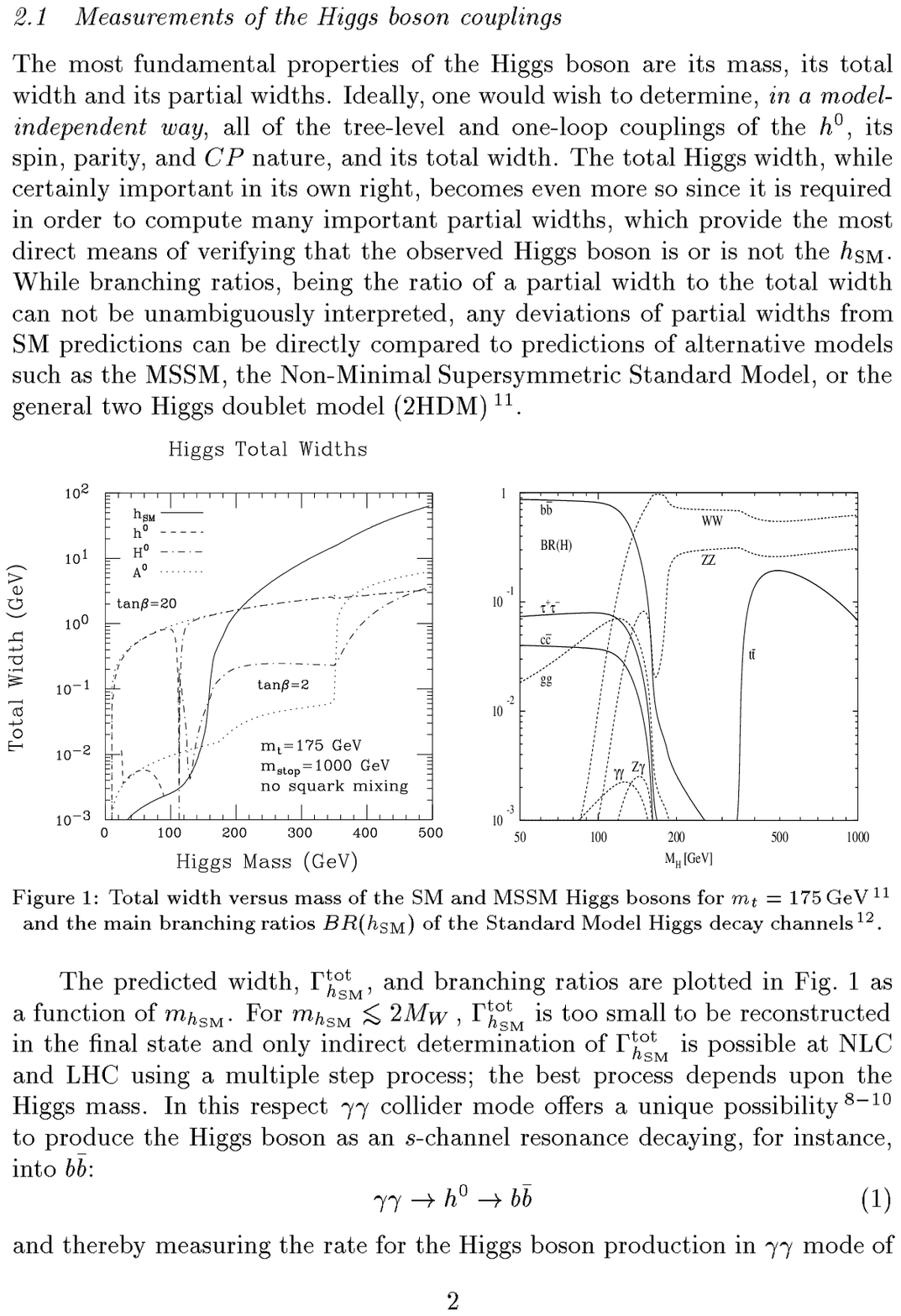}
}
\end{minipage}

\vspace{0.3cm}

\caption{Some of the SM Higgs branching ratios as a function
of its mass~\cite{BAIL94,DJOUADI}. On the left figure~\cite{BAIL94},  
the total width and the $\gamma\gamma$ width (thick lines) are also shown.}
\label{hbranch}
\end{figure}

Search and study of the Higgs boson, the key particle of the Standard
model, is one of the primary goals of linear colliders. The LEP-2 will
put the lower limit on its mass ot about 95 GeV.  Indirectly, from
radiative corrections, it follows that the Higgs mass (if the Standard
Model is correct) is $140^{+150}_{-80}$ GeV~\cite{BLONDEL}. The MSSM
predicts a mass of the lightest neutral Higgs less of than 130
GeV. So, the mass of the Higgs most probably lies in the region of
100$<M_H<$300 GeV.  Some parameters of the SM Higgs (total width, \GG\
width, and main branching ratios) are presented in
fig. \ref{hbranch}$\;$~\cite{BAIL94,DJOUADI}.

  Higgs production in \GG\ collisions has been considered in many
papers. Unfortunately, different authors have used their own definitions
of the \GG\ luminosity which often led  to underestimation of the
expected number of events for a given integrated \GG\ luminosity. In the
present paper, we will directly compare  the Higgs production cross
sections in \GG\ and \EPEM\ collisions. I would like also to pay  
attention to some additional possibilities for Higgs studies at photon
colliders connected with a very sharp edge in the \GG\ luminosity spectra.


 The total Higgs width at masses below 400 GeV is much smaller than
the characteristic width of \GG\ luminosity spectra (FWHM $\sim 10-15 \%$
), therefore the production rate is proportional to $d\LGG/W_{\GG}$:
\begin{equation}
\dot{N}_{\GG\ \to H}
=\LGG \times \frac{d\LGG M_H}{d\WGG \LGG}\frac{4\pi^2\Gamma_{\GG}
(1+\lambda_1 \lambda_2)}{M^3_H}\equiv \LGG \times \sigma^{eff}.
\end{equation}
where $\lambda_i$ is the photon helicity.
  Below we  assume that the Higgs search and study is done utilizing the high
energy peak of the \GG\ luminosity energy spectrum.
 
\begin{figure}[!htb]

\vspace{-1cm}

\hspace*{0cm}\begin{minipage}[b]{0.45\linewidth}
\centering
\vspace*{-0.cm} 
\hspace*{-1.cm} \epsfig{file=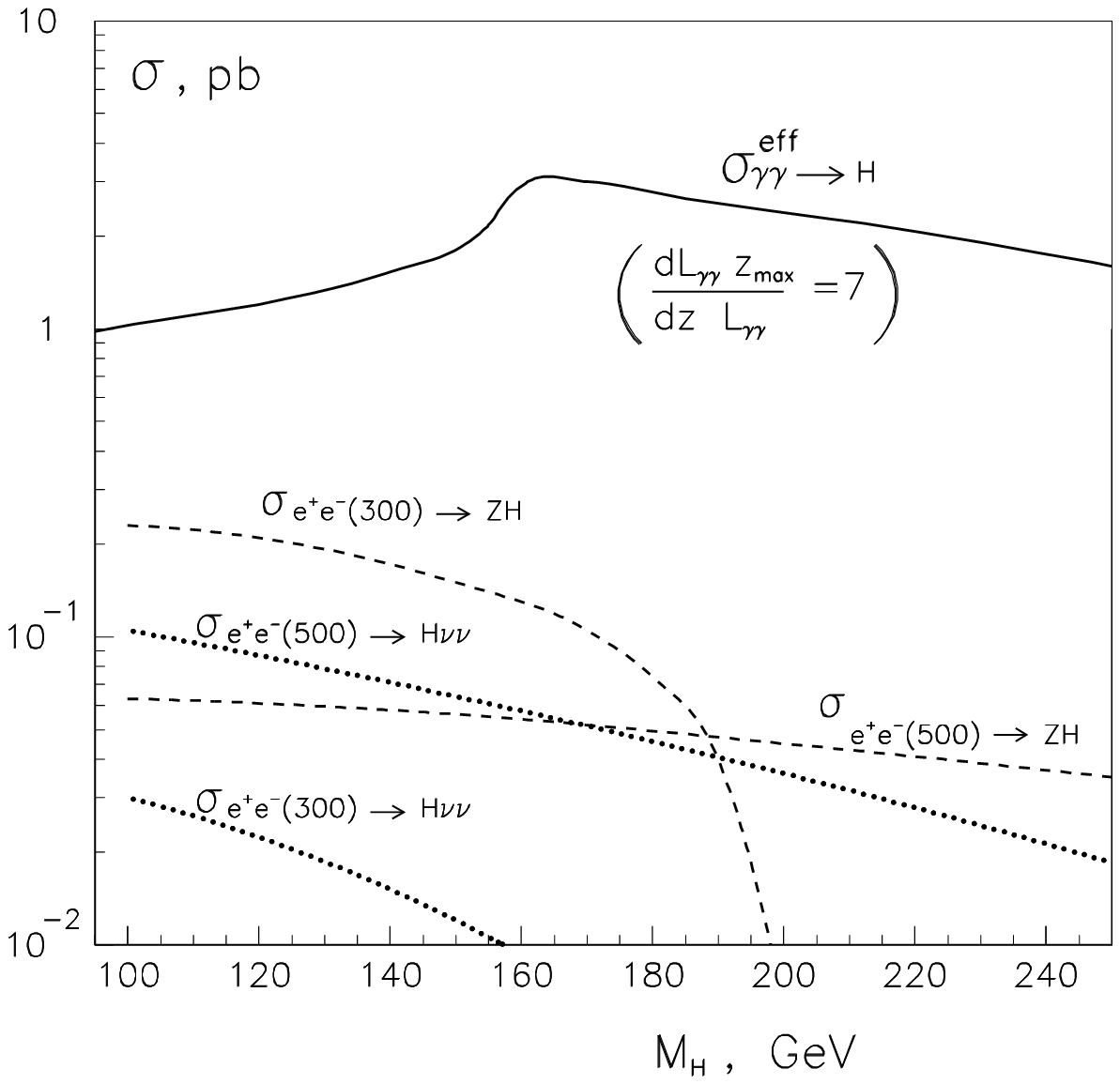,width=8.cm,angle=0} 

\vspace{-0.4cm}

\caption{Cross sections for the Standard model Higgs in \GG\ and
 \EPEM\ collisions.}
\label{cross}
\end{minipage}%
\hspace*{1.1cm} \begin{minipage}[b]{0.45\linewidth}
\centering
\vspace*{0.cm} 
\hspace*{-1.4cm} \epsfig{file=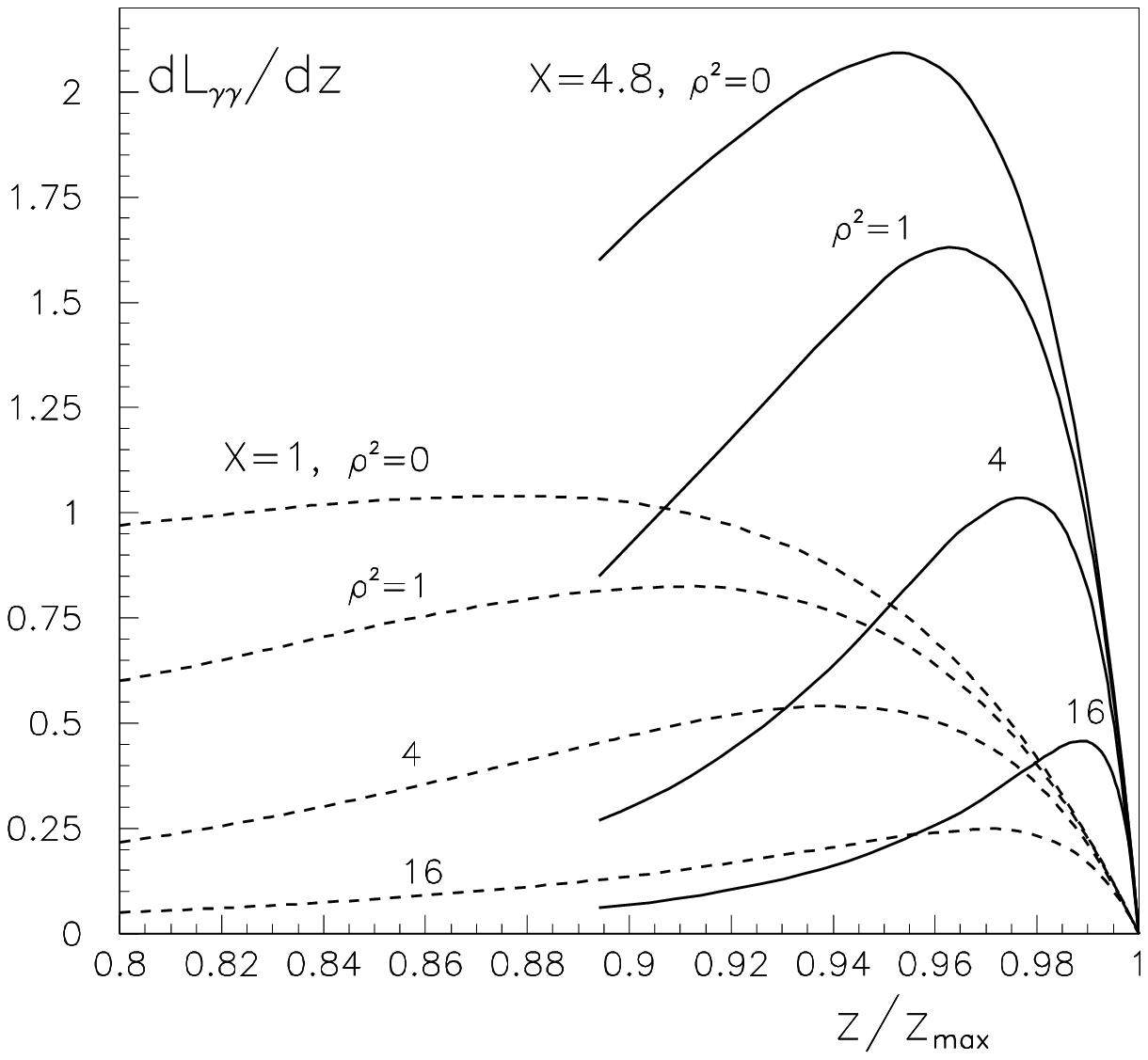,width=8cm,angle=0}

\vspace{-0.4cm}

\caption{ Shape of the \GG\ luminosity spectrum near the high energy edge.
Here $z=W_{\GG}/2E_0.$}
\label{edge}
\end{minipage}
\end{figure}

The effective cross section for $(d\LGG/d\WGG) (M_H/\LGG)=7$ and 
$1+\lambda_1 \lambda_2=2$ is presented in fig.~\ref{cross}. 
Note that here \LGG\
is defined as the \GG\ luminosity at the high energy luminosity peak
($z=\WGG/2E_e>0.65$ for $x=4.8$), thus ignoring the lower energy part
of the luminosity spectrum, 
which is much less valuable for experiment.\footnote{It 
is also possible to search for the Higgs at a constant
collider energy utilizing a flat part of the luminosity spectrum
instead of energy scanning. However, in this case $d\LGG/d\WGG\ $ is
lower (no peak and a wider energy range) and we have much larger
backgrounds ($\gamma g \to b\bar{b}$, etc), worse polarization degree
(partially due to contribution of multiple Compton scattering and
unpolarized beamstrahlung photons) and, consequently, worse
suppression of backgrounds; therefore, the required total integrated
luminosity will be only larger than in the method of scanning (this
statement should be checked more carefully). If the Higgs mass is
already known, then, obviously, one has to work at $\WGG_{max}\sim
M_H$.} 
 This luminosity is approximately equal to
$0.25k^2\LEE(geom)$.  For comparison, in the same figure the cross
sections of the Higgs production in \EPEM\ collisions is shown. We see
that for $M_H=$ 100--250 GeV the effective cross section in \GG\
collisions is larger than that in \EPEM\ collisions by a factor of
about 5--30. This interesting fact has never been emphasized.



How do we study the Higgs in \GG\ collisions? One obvious way is
searching for a peak in the invariant mass distribution  measured by
the detector. This is the only method  utilizing the broad
luminosity spectrum. The mass resolution in this method can be somewhat better
than the width of the high energy luminosity peak. 

Another method is the energy scanning where  we can use some important
features of the \GG\ luminosity distribution: quite narrow width of
high energy peak and a {\it very sharp edge of luminosity distribution}
which is much narrower than the width of the luminosity peak and the detector
resolution, see fig. \ref{edge}. For $x=4.8$ the differential
luminosity ($dL/dz$) reaches  half of its maximum  at $\Delta
z=z_{max}-z=0.8\%$ (at $\rho = b/\gamma a_e=1$).
During the scanning, we will observe a sharp increase in the visible cross
section when the maximum energy of the \GG\ collider reaches $M_H$ (if
the Higgs is a very narrow resonance). Observation of a step at the same
energy for different decay modes significantly increases confidence of
results, especially for modes with small branchings or with
difficulties in reconstruction. For example, detection of the
$H\to\tau\tau$ decay is a difficult task due to undetected neutrinos. The most
reliable way to see this channel is to observe a step in visible
cross section for events with the following selection criteria: two collinear
low-multiplicity jets with unbalanced transverse momentum and with the energy
of one jet not far from the maximum photon energy.

 The total number of events for 10 fb$^{-1}$ of integrated \GG\
luminosity (as it was defined above) for the case when the peak of the
luminosity spectrum sits at the Higgs boson mass is 13000 (18000) for
$M_H$ = 130 (150) GeV respectively (follows from cross section given
in fig. \ref{cross}). That is a lot! Due to energy scanning, the
statistics will be somewhat smaller, about $N \approx
0.5N_{peak}$. The number of events for various decay modes (without
any cuts) is given in table \ref{tabl2}.

\begin{table}[htb]
\caption{The number of  Higgs events in various decay modes (no
cuts) for $M_{H}$= 130 and 150 GeV and 10 fb$^{-1}$ of integrated \GG\
luminosity.}
\begin{center}
\begin{tabular}{l|cccccccc} \hline
       Mode & $b\bar{b}$ &WW$^*$&ZZ$^*$&$\tau\tau$&$c\bar{c}$&gg&
\GG\ &$\gamma$ Z\\ 
\hline\hline
Br $M_H(130)$ &0.52&0.29&0.037&0.055&0.025&0.063&0.0022&0.0019\\ \hline

events&3400&1900&240&350&160&410&15&12 \\ \hline\hline

Br $M_H(150)$ &0.18&0.67&0.083&0.019&0.008&0.03&0.0014&0.002\\ \hline
events&1600&6000&750&171&74&260&13&22 \\ \hline\hline
\end{tabular}
\end{center}
\label{tabl2}
\end{table}

  The Higgs production at photon colliders and various backgrounds have been
  studied in many papers for various decay modes:
  $b\bar{b}\;$
\cite{BARKLOW}-\cite{OHGAKI97};
 ZZ~\cite{BBC1,GUNION93,JIKIAZZ};
  WW~\cite{BBC1,GUNION93,MORRIS,GINZ97}. 

Below $M_H < 150$ GeV, the SM Higgs decays mainly into $b\bar{b}$
pairs. The main backgrounds here are the QED processes $\GG\ \to
b\bar{b},\; c\bar{c}$, which can be suppressed using  vertex detectors and
polarized photon beams ($\sigma_{\gamma\gamma \to q\bar{q}} \propto
1-\lambda_1\lambda_2$, while $\sigma_{\gamma\gamma\to H} \propto
1+\lambda_1\lambda_2$). The remaining background is small.

  For $M_H>190$ GeV, the Higgs can be observed in the best way in the ZZ
  decay mode. The main background here is the process $\gamma\gamma \to
  WW$, which can be suppressed by requiring that at least one of the Z's  be
  detected in $l^+l^-$ decay modes. In this channel, the SM Higgs can
  be observed in the range $M_H \sim$ 120 -- 350 GeV. At higher masses,
  the Higgs signal will be much smaller than irreducible background
  $\gamma\gamma \to ZZ$~\cite{JIKIAZZ}.

In the region $M_H > 140$ GeV, the SM Higgs decays mainly
into $WW$ (or $WW^*$) pairs. The main problem here is the background from
$\gamma\gamma \to WW$  with large cross section. Fortunately, this
background is very small for $M_H <2M_W$~\cite{GUNION93,MORRIS,GINZ97}, 
so that  Higgs with $100 <M_H < 160$ GeV
can be detected in this channel practically without backgrounds.  

\begin{figure}[thb]
\hspace*{1cm}\begin{minipage}[b]{0.45\linewidth}
\centering
\hspace*{-1.6cm}
\scalebox{1.05}[1.05]{
\includegraphics[width=7cm, angle=0, bb=104 451 471 772,clip] {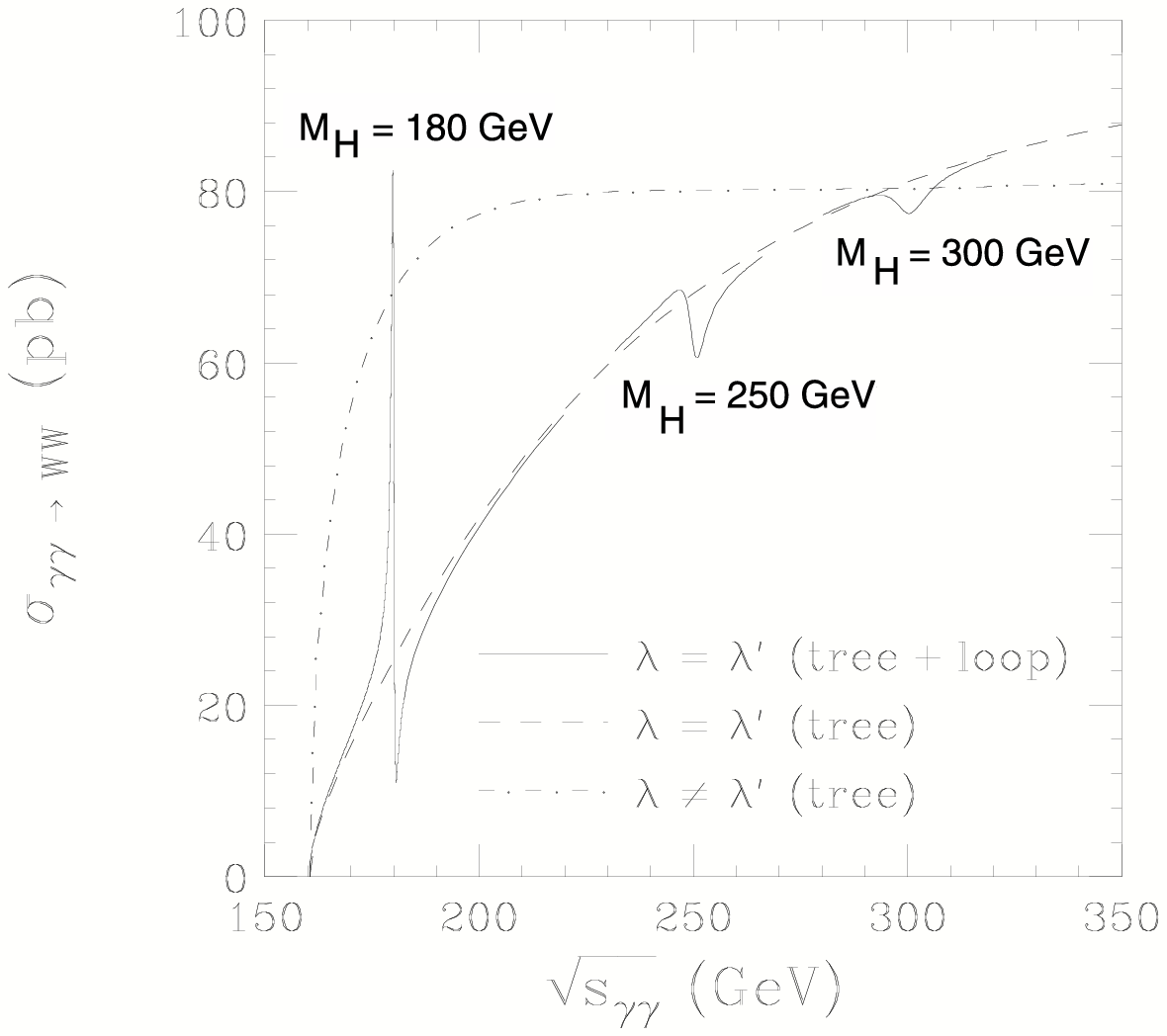}}
\end{minipage}%
\hspace*{-0.cm} \vspace*{-0.cm} \begin{minipage}[b]{0.45\linewidth}
\centering
\vspace*{-0.3cm}
\includegraphics[width=6cm, angle=0, bb=104 229 504 657,clip] {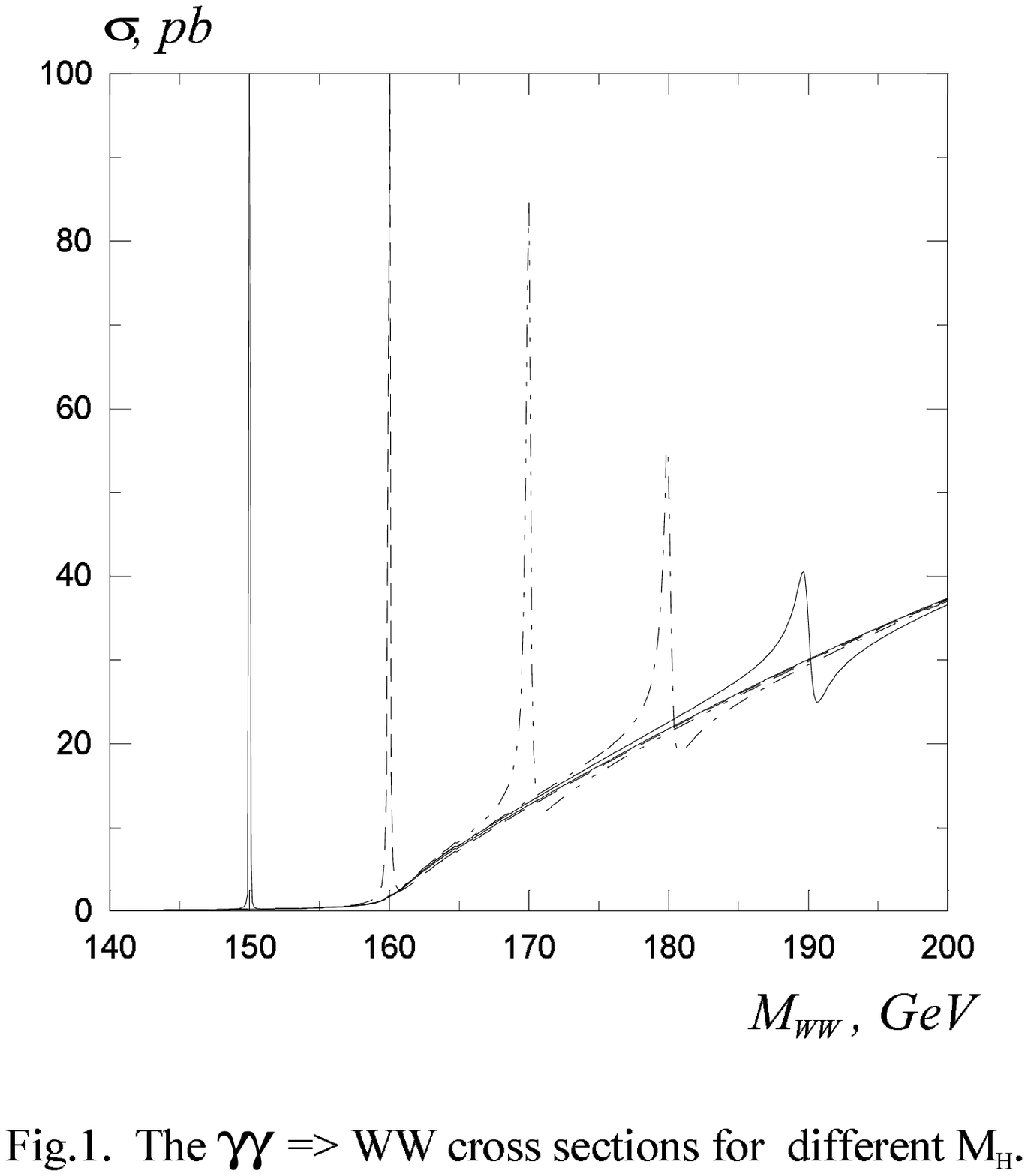}
\end{minipage}

\vspace{0.2cm}

\caption{The $\gamma\gamma\to WW$ cross section for various Higgs masses, 
see comments in the text}
\label{WW}
\end{figure}

 In the region $M_H > 160$ GeV, the interference between the Higgs
signal and the background becomes important~\cite{MORRIS,GINZ97}.  The cross
section of $\GG\ \to $ WW for various Higgs masses is shown in
fig.~\ref{WW} (the left figure is taken from ref.\cite{MORRIS}, the right one
from ref.\cite{GINZ97}) These cross sections correspond to monochromatic photon
beams. In a real situation the photon beam energy spread and the
detector resolution are much larger than the Higgs width, so that the
resonance structures is much broader but nevertheless observable for
not too large $M_H$. For $M_H>2M_W$, there are four additional
constraints in the 4-jet decay mode: two come from the zero transverse
momentum of the system, and two come from the jet-jet mass equal to
$M_W$. Using of these constraints should improve the WW invariant mass
resolution.

Much better sensitivity to narrow structures can be obtained utilizing
the sharp edge of the \GG\ luminosity spectrum. Using all these
methods, one can see the Higgs in $\GG\ \to WW$ decay mode from
$M_H\sim 100$ GeV up to $M_H \sim$ 200--250 GeV or even the higher.

Detection of the Higgs in $\tau\tau,cc,gg,\gamma\gamma,\gamma Z$ decays are
also possible at photon colliders. In \GG\ and $\gamma Z$ modes, the
background is rather small~\cite{JIKIAgg,JIKIAgZ}. Conclusions on
perspectives of Higgs detection  in these modes and required
luminosities can be made only after a detail simulation.

  Photon colliders also allow to measure the Higgs mass with a high
  precision. The very sharp edge of the luminosity spectrum and high
  statistics make it possible to obtain a much better  accuracy  than 
  in \EPEM\ or pp collisions. In this measurement, it is important to
  remember that nonlinear effects in the conversion region can shift
  the maximum energy of Compton photons and change the shape of the
  luminosity spectrum~\cite{TEL90,TEL95}. To reduce this effect, the
  parameter $\xi^2$ characterizing nonlinearity should be kept small
  enough. The shape of the luminosity spectrum near $z_{max}$ should
  be measured very precisely using the processes $\GG\ \to \mu\mu\;
  (\EPEM)$.

As we have seen, the Higgs can be very successfully studied in \GG\
collisions for $M_H$ = 100--350 GeV,  may be even better than
in \EPEM\ collisions.

\section{Luminosity}
\subsection{Current projects}

 Due to the absence of beamstrahlung, beams in \GG\ collisions can
have much smaller horizontal beam size than  in \EPEM\ collisions,
thus the beta functions at the interaction point can be made as
small as possible (some restrictions are imposed by the Oide effect
connected with chromatic aberrations due to synchrotron radiation in
the final quads).  However, even after optimization of the final focusing
system the attainable \GG\ luminosity in curent LC projects is
determined by the attainable ``geometric'' ee--luminosity.

 The results of simulations for different projects are the
   following. For the ``nominal'' beam parameters (the same as in
   \EPEM\ collisions) and the optimum final focus system the
   luminosity $\LGG(z>0.65) \sim (0.8/1.2/0.7)\times10^{33}$ \CMS\ for
   NLC/TESLA/SBLC, or by about factor 5 smaller than 
   \EPEM\ luminosity.

 Obviously, this is not a fundamental limit. Below we will discuss what
 the limit really is and how to approach  it.

\subsection{Ultimate luminosity}

  The only collision effect restricting \GG\ luminosity at photon
  colliders is the coherent pair creation which leads to the
  conversion of a high energy photon into \EPEM\ pair in the field of
  the opposing electron beam~\cite{CHEN,TEL90,TEL95}. There are
  three ways to avoid this effect: a) use flat beams; b) deflect the
  electron beam after conversion at a sufficiently large distance from
  the IP; c) under certain conditions (low beam energy, long bunches)
  the beam field at the IP is below the critical one due to the
  repulsion of electron beams~\cite{TELSH}.  The problem of ultimate
  luminosities for different beam parameters and energies was analyzed
  recently in ref.\cite{TSB2} analytically and by simulation. The  resume
  is  following.

\begin{figure}[!hb]
\centering
\vspace{3mm}
\epsfig{file=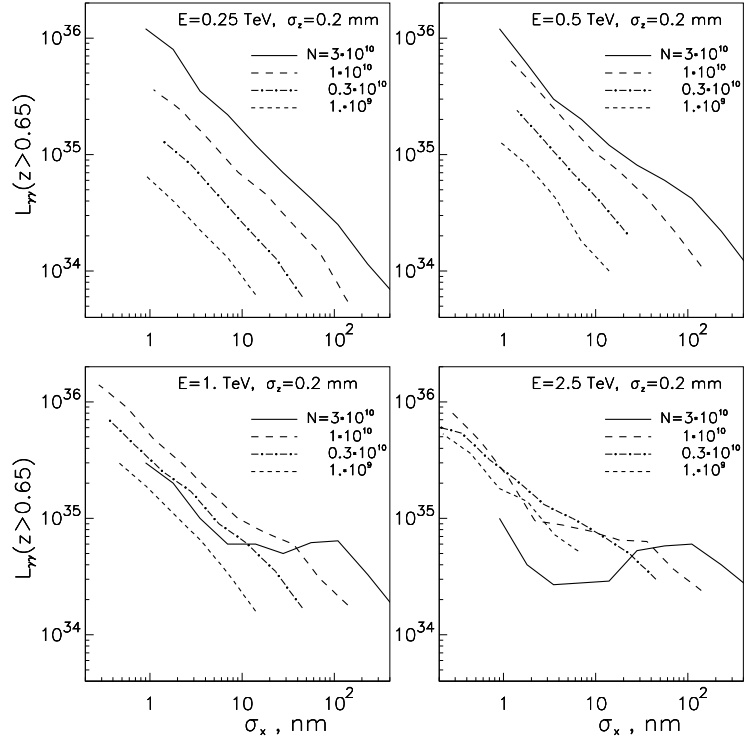,width=10cm,bb= 32 50 550 540 }
\caption{Dependence of the \GG\ luminosity on the horizontal beam size
for $\sigma_z = 0.2$ \MM, see comments in the text.}
\label{sb3}
\end{figure}

  The maximum luminosity is attained when the conversion point is
  situated as close as possible to the IP, at $b \approx
  3\sigma_z+0.04E[\TEV]$ cm (here the second term is equal to the
  minimum length of the conversion region). In this case, the vertical
  radius of the photon beam at the IP is also minimal: $a_{\gamma}\sim
  b/\gamma$ (assuming that the vertical size of the electron beam is
  even smaller). The optimal horizontal beam size ($\sigma_x$) depends
  on the beam energy, the number of particles in a bunch and the bunch
  length.  The dependence of the \GG\ luminosity on $\sigma_x$ for
  various energies and numbers of particles per bunch is shown in
  fig. \ref{sb3}. The bunch length is fixed at 0.2 mm. The collision
  rate is calculated from the total beam power, which is equal to
  15E[TeV] MW (close to that in current projects). From the
  fig. \ref{sb3} we see that at low energies and small numbers of
  particles, the luminosity curves follow their natural behavior $L
  \propto 1/\sigma_x$, while at high energies and large numbers of
  particles per bunch the curves make a zigzag which is explained by
  $\gamma \to$ \EPEM\ conversion in the field of the opposing beam.

   What is so remarkable about these results? First of all, the maximum
   attainable luminosities are huge, above $10^{35}$ \CMS. At low
   energies, there is no coherent pair creation, even for a  very small
   $\sigma_x$'s, when the field in the beam is much higher than the
   critical field $B_{cr}=\alpha e/\gamma r^2_e$. This is explained by
   the fact that during the collision the beams are repulsing each other
   so that the field on the beam axis (which affects  high energy
   photons) is below the critical field. It means that the
   $\gamma\gamma$ luminosity is simply proportional to the geometric
   electron-electron luminosity (approximately $\LGG(z>0.65)\sim 0.1
   L_{ee}$) for $\sigma_x,\sigma_y < b/\gamma \sim 3\sigma_z/\gamma +
   0.2$ nm.  For  energies 2E$ <$ 2 TeV, which are in reach of the next
   generation of linear colliders, the luminosity limit is much higher
   than it is usually required  and much higher than in
   \EPEM\ collisions (especially at low energies).

  One of the main problems in obtaining such high \GG\ luminosity is
generation of electron beams with small emittances in both horizontal
and vertical directions. In the following sections we will discuss
various approaches to solving this problems.

\subsection{Ways to  higher luminosities}

There are several possibilities for increasing  luminosity. 

1) Reduction of the horizontal emittance by optimizing the damping
rings. For example, at the TESLA, a decrease of \ENX\ by a factor of
3.5 leads to an increase in \LGG\ up to $3\times 10^{33}$
\CMS. However, it seems that this way is quite difficult.

2) One can use low emittance RF-photoguns instead of damping
rings. Unfortunately, even with the best photoguns the luminosity will
be somewhat lower than that with damping rings. However, there is one
possible solution. The normalized emittance in photoguns is
approximately proportional to the number of particles in the electron
bunch. It seems possible to merge (using some difference in energies)
many ($N_g \sim 5-10$) low current beams with low emittances to one
high current beam with the same transverse emittance. This gives us a
gain in luminosity more than by a factor of $N_g$ in comparison with a
single photogun (``more'' because the lower emittance allows smaller
beta functions due to the Oide effect).  Estimations show that with
ten photoguns, one can achieve \LGG\ $\sim L_{\EPEM}$ in all
considered projects.  There is only one ``small'' problem: such
RF-guns with polarized electrons do not exist yet, though there are no
visible fundamental problems~\cite{CLENDENIN}. Also there is the
``thermal'' limit on emittance in photoguns, which puts some
restrictions on this method. In the AsGa photoguns with polarized
electrons, this limit should be much smaller than for metal
photocathodes~\cite{CLENDENIN}.

3) If one has flat electron beams (\ENX $>>$ \ENY) from some injector
(a damping ring or, better, an RF-gun), then it is possible, in principle, to
make exchange of the horizonlal and longitudinal emittances using
an off-axis RF cavity (some head-tail kick arizing here can be removed by
additional RF cavities rotated by 90$^0$. This method was never tested
or even discussed. I do not know why. This method is especially
promising for beams from RF-guns, where very small energy spread
allows both horizontal and longitudinal beam compression.
This method as well as the previous one should be  studied more carefully.

4) For a considerable step up in luminosity,
beams with much lower emittances are required. That needs
development of new approaches, such as {\it laser cooling}~\cite{TSB1}
(see next subsection).  Potentially, this method allows to attain a
geometric luminosity by two orders higher than that achievable by
other known methods.

\subsection{Laser cooling}

  Recently~\cite{TSB1}, a new method of beam production with small
emittance was considered --- laser cooling of electron beams --- which
allows, in principle, to reach $\LGG \geq 10^{35}$ \CMS.

The idea of laser cooling of electron beams is very simple.  During a
collision with optical laser photons (in the case of strong fields it
is more appropriate to consider the interaction of an electron with an
electromagnetic wave) the transverse distribution of electrons
($\sigma_i$) remains almost the same. Also, the angular spread
($\sigma_i^{\prime}$) is almost constant, because electrons loss  momenta
 almost along their
trajectory (photons follow the initial electron trajectory with a
small additional spread). So, the emittance $\EI = \SI \SIP$ remains
almost unchanged. At the same time, the electron energy decreases from
$E_0$ down to $E$. This means that the transverse normalized
emittances have decreased: $ \EN = \gamma \E = \EN_0(E/E_0)$.  One can
reaccelerate the electrons up to the initial energy and repeat the
procedure. Then after N stages of cooling $ \EN /\EN _0 = (E/E_0)^N$
(if \EN\ is far from its limit).

 Some possible set of parameters for  laser cooling is: $E_0 = 4.5$ GeV,
$l_e=0.2 $ mm, $\lambda = 0.5$ \MKM, flash energy $A \sim 10 $ J.  The
final electron bunch will have an energy of 0.45 \GEV\ with an energy
spread $\sigma_E/E \sim 13 \%$, the normalized emittances \ENX,\ENY\
are reduced by a factor of 10.  A two-stage system with the same parameters
reduces the emittances by a factor of 100. The limit on the final
emittance is $\ENX\ \sim\ENY\ \sim2\times 10^{-9}\;$ m~rad at
$\beta_i= 1\; \MM$.  For comparison, in the TESLA (NLC)
project the damping rings have $\ENX\ =14(3)\times 10^{-6}\;$ m~rad,
$\ENY\ =25(3)\times 10^{-8}\;$ m~rad. 

This method requires a laser system even more powerful than that for e
$\to \gamma$ conversion. However, all the requirements are reasonable,
taking into account the fast progress of laser technique and time
plans of linear colliders. Multiple use of the laser bunch (some
optical resonator) can  considerably reduce the required  average laser power.
This method can be tested already now at a low repeatition rate.

\section{Acknowledgements}

 I would like to thank Clem Heusch for a nice Workshop, which was
 one of the important steps towards \EPEM, ee, \GE, \GG\ colliders.

\end{document}